# Magnetic-field-controlled negative differential conductance in graphene *npn* junction resonators

Si-Yu Li[1], Haiwen Liu[1], Jia-Bin Qiao[1], Hua Jiang[2,*], and Lin He[1,*]

[1]Center for Advanced Quantum Studies, Department of Physics, Beijing Normal University, Beijing, 100875, People's Republic of China

[2]College of Physics, Optoelectronics and Energy and Institute for Advanced Study, Soochow University, Suzhou, 215006, People's Republic of China

Correspondence and requests for materials should be addressed to H.J. (e-mail: jianghuaphy@suda.edu.cn), and L.H. (e-mail: helin@bnu.edu.cn).

**Negative differential conductance (NDC), characterized by the decreasing current with increasing voltage, has attracted continuous attention for its various novel applications. The NDC typically exists in a certain range of bias voltages for a selected system and controlling the regions of NDC in curves of current versus voltage (*I-V*) is experimentally challenging. Here, we demonstrate an unusual magnetic-field-controlled NDC in graphene *npn* junction resonators. The magnetic field not only can switch on and off the NDC, but also can continuously tune the regions of the NDC in the *I-V* curves. In the graphene *npn* junction resonators, magnetic fields generate sharp and pronounced Landau-level peaks with the help of the Klein tunneling of massless Dirac fermions. A tip of scanning tunneling microscope induces a relatively shift of the Landau levels in graphene beneath the tip. Tunneling between the misaligned Landau levels results in the magnetic-field-controlled NDC that may have potential applications for future graphene-based technology.**

The Klein tunneling, as one of the most fantastic properties in graphene, describes an unusual anisotropic transmission of the massless Dirac fermions across *p-n* junctions in graphene[1-6]. Because of this unique property, *p-n* junctions in graphene can be used to manipulate electron refraction and transmission like photons and, therefore, they become essential building blocks in quantum electron optics based on graphene[7-15]. In the past few years, *npn* junctions have been introduced in continuous graphene sheets as electron resonators and the massless Dirac fermions of graphene are temporarily trapped to form quasibound states inside the resonators because of the Klein tunneling[16-21]. By introducing a small perpendicular magnetic field on the order of 0.1 T, it was demonstrated very recently that the Berry phase can be switched on and off and, consequently, there is a sudden and large change in the energy of angular-momentum states in graphene *npn* junction resonators[22]. Such a result indicates that many exotic phenomena are expected to be uncovered in the graphene *npn* junction resonators under magnetic fields.

In this work, we report an unusual phenomenon, magnetic-field-controlled negative differential conductance (NDC), in the graphene *npn* junction resonators in the presence of large magnetic fields. The NDC, which is the phenomenon of decreasing current with increasing voltage, usually exists in certain regions of bias voltages in the current-voltage (*I-V*) curves[23-32]. For a selected system, it is experimentally challenging to tune the regions of bias voltages where the NDC occurs. In our experiment, a large perpendicular magnetic field generates quite sharp and pronounced Landau-level peaks with the help of electron confinement of the graphene *npn* junction resonators. The tip of scanning tunneling microscope (STM) induces a local potential in graphene beneath the tip, which results in a relatively shift of the Landau levels (LLs). Tunneling between the misaligned LLs leads to a strong NDC accompanying the LLs in scanning tunneling spectroscopy (STS) spectra. Since the magnetic field controls both the Landau quantization and the energy positions of the LLs in graphene, therefore, we can use the magnetic field not only to switch on and off the NDC but also to continuously tune the regions of the NDC in the spectra.

In our experiment, graphene systems were synthesized by chemical vapor deposition

(CVD) on Rh foils (see Supplementary Fig. S1 for details of sample preparation) and we have demonstrated previously that the adjacent graphene sheets grown on Rh foils have a strong twisting tendency[33-35]. Usually, the topmost graphene monolayer electronically decouples from underlying graphene sheets and behaves as free-standing graphene monolayer due to the existence of large twist angles[34-37]. Figure 1 shows representative STM images and STS ($dI/dV$-$V$) spectra of a graphene *npn* junction studied in this work (see Supplementary Fig. S2 for a large STM image of the graphene on Rh foil and Supplementary Fig. S3 for more STS spectra). According to the high-field spectra shown in Fig. 1c, two important results about the sample can be obtained. First, the topmost graphene monolayer electronically decouples from the supporting substrate. We observed well-defined Landau quantization of massless Dirac fermions[35-39]: the observed Landau level (LL) energies depend linearly on the square root of both level index *n* and magnetic field *B*, as expected to be observed in free-standing graphene monolayer (see Supplementary Fig. S4 for detail of analysis). Such a result demonstrated explicitly that the topmost graphene monolayer behaves as free-standing graphene monolayer. Second, the electronic doping in the topmost graphene monolayer is inhomogeneous because different charge transfers from the substrate[18,19,40,41]. The zero ($n = 0$) LL directly reflect the charge neutrality point (Dirac point) of the graphene monolayer. According to the result in Fig. 1c, the center of the studied region (Fig. 1a) is *p*-doped (the Dirac point is about 30 meV above the Fermi level), and the rest is *n*-doped (the Dirac point is about 20 meV below the Fermi level), as schematically shown in Fig. 1d. Based on the high-field STS measurements, we can explicitly identify boundaries of the *p*-doped region in the graphene *npn* junction (the dashed lines in Fig. 1a).

Although the potential barrier across the *npn* junction is only about 50 meV (Fig. 1c), it results in quite different electronic properties (STS spectra) between the *p*-doped region and the *n*-doped region. One obvious difference between the STS spectra recorded in the two different regions is the intensity asymmetry between the states above the Dirac point and that below the Dirac point. In the *p*-doped region, the intensity of states below the Dirac point is much stronger than that above the Dirac point in the spectra recorded both in zero magnetic field (Fig. 1b) and in the presence of high magnetic fields (Fig. 1c). Whereas, we observed the opposite case in the *n*-doped region. Such a result is reasonable because that charge carriers below the Dirac point can be temporarily trapped inside the *p*-doped region to form quasibound states in the graphene *npn* junction, which could enhance the intensity of states below the Dirac point in the *p*-doped region[16-21]. Additionally, such an effect may become more remarkable due to the Landau quantization effect. Then the lifetime of the quasiparticles is much increased and, consequently, we observed sharp and pronounced LL peaks below the Dirac point in the *p*-doped region, as shown in Fig. 1c.

The primary difference between the STS spectra recorded in the two different regions is the emergence of NDC in the *p*-doped region, whereas not in the *n*-doped region (Fig. 1c). To clearly show this primary difference, we plot two representative STS spectra recorded at 6 T in the two different regions in Fig. 2a. Obviously, there are four regions of strong NDC accompanying four sharp and pronounced LLs ($n$ = 0, -1, -2, -3) in the spectrum recorded in the *p*-doped region. The characteristic decreasing current with increasing bias voltage of the NDC could also be clearly identified in the *I-V* curve recorded in the *p*-doped region (the upper panel of Fig. 2b). However, we cannot detect any signature of NDC in the spectra recorded in the *n*-doped region. In Fig. 2c, we show an energy-fixed STS mapping recorded at a bias voltage in a region that the NDC occurs, which directly reflects the positions where we can observe the NDC in real space (see Supplementary Fig. S5 for more STS mapping at other bias voltage). Such a measurement demonstrated explicitly that the observed NDC is closely related to the *p*-doped region in the graphene *npn* junction. Similar result was also obtained in several different graphene *npn* junctions in our experiment (see Supplementary Fig. S6 for more experimental data), indicating that the NDC is a general and robust feature in the *p*-doped regions of the graphene *npn* junctions in the presence of high magnetic fields.

To further explore the origin of the NDC, we measured Landau quantization in the *p*-doped region with different magnetic fields (see Supplementary Fig. S4 for the STS spectra). The regions of the NDC always accompany the sharp and pronounced LLs of the STS spectra in the *p*-doped region. Figure 3a shows the evolution of the series of

NDC as a function of magnetic field. Each panel of the indicated magnetic field consists of 10 STS spectra acquired at different positions of the *p*-doped region in Fig. 1a. The STS intensity is shown in a color scale and we only show the tunneling conductance below zero to emphasize the NDC. Obviously, the spectra recorded at different positions are reproducible and the NDC can be controlled by the magnetic field. Actually, the series of the NDC show the same dependence as the LLs on the magnetic field, strongly suggesting those LLs as source of the observed NDC. In our experiment, only the sharp and pronounced LLs lead to the NDC (Fig. 1 and Fig. 2). To clearly show the close relationship between the sharp and pronounced LLs and the NDC, we summarize the ratios *h*/*d* for different LLs measured at different magnetic fields in the *p*-doped region and the *n*-doped region in Fig. 3b. Here *h* is the height of the LLs and *d* is the full width at half maximum of the LLs in the STS spectra, as defined in Fig. 2a. Obviously, the NDC can only be observed accompanying the LLs with large values of the ratio *h*/*d* (above the dashed line in Fig. 3b), *i.e.*, the LLs that are sharp and pronounced in the spectra. As a consequence, the NDC can only be observed in the *p*-doped region of the graphene *npn* junction.

The vital role of the sharp and pronounced LLs in the emergence of the NDC is also confirmed in our theoretical calculations. Our analysis and simulation demonstrate that the observed NDC arises from tunneling between slightly misaligned LLs with large values of the ratio *h*/*d*. In our experiment, the STM tip was not only used to probe the local density-of-state (LDOS) of the graphene, but also as source of electrostatic potential to gate electronic states of the sample beneath the tip[16,20,22], as schematically shown in Fig. 4a. In the presence of perpendicular magnetic fields, the Hamiltonian of a graphene region locally beneath the STM tip can be written as $H_{QD} = \sum_{i=-N}^{N}(\epsilon_i + \alpha eV)c_i^+ c_i$. Here, the $\epsilon_i$ represents the energies of the LLs in the *p*-doped region, $\alpha eV$ describes the energy shift of the LLs induced by the bias of the STM tip *V* (the factor $\alpha$ is the tip lever arm), and $c_i^+$ and $c_i$ are the creation and annihilation operators (see Supplementary materials for detail of calculation). The electrons inject from the STM tip into the graphene region beneath the tip should tunnel to the surrounding bulk LLs. In such a case, the variation of the current of the STM tip with

respect to the change of the voltage bias should be expressed as

$$I(V+dV) - I(V) = \frac{e}{h}\left[\int_{E_F}^{E_F+e(V+dV)} T(E,V+dV)\,dE - \int_{E_F}^{E_F+eV} T(E,V)\,dE\right]$$

$$= \frac{e^2}{h}T(E_F+eV)dV + \frac{e}{h}dV\int_{E_F}^{E_F+eV}\frac{\partial T}{\partial V}dE. \quad (1)$$

Here $T(E,V)$ is the transmission coefficient from STM tip to the sample. Then we obtained the differential conductance $G = dI/dV$ as

$$G = \frac{e^2}{h}T(E_F+eV) + \frac{e}{h}\int_{E_F}^{E_F+eV}\frac{\partial T}{\partial V}dE. \quad (2)$$

The first term in Eq. (2) is the conventional differential tunneling conductance, which is proportional to the realistic LDOS of the sample. Therefore, we still can observe LDOS of the LLs peaks in the STS spectra, as shown in Figs. 1-3. The second term involve the variations of the transmission coefficient in the energy region $[E_F, E_F+eV]$. Due to the formation of sharp and pronounced LLs in high magnetic fields, the LDOS changes sharply by varying the energy $E$. When the LLs in the graphene region beneath the tip and that of surrounding bulk become misaligned with increasing $V$ (as schematically shown in Fig. 4b), the transmission coefficient $T$ decreases in the region $[E_F, E_F+eV]$. As a consequence, the second term of Eq. (2) becomes negative, which could lead to a strong NDC in the spectra when $T(E_F+eV)$ in the first term approaches zero. Figure 4c and Fig. 4d show the representative *I-V* curve and *dI/dV-V* spectrum calculated according to the Eq. (1) and Eq. (2), respectively. In the simulation, the height and the full width at half maximum of the LLs obtained in our experimental STS spectra are taking into account (see Supplementary for details of model, numerical methods and parameters). Obviously, the main features of our experimental results are well reproduced in our calculation, as shown in Fig. 4c-4e. The existence of sharp and pronounced LLs and the tip-induced energy shift of the LLs beneath the tip are the two key reasons for the emergence of the NDC. In our experiment, we can use the magnetic field to control both the Landau quantization and the energy positions of the LLs in graphene monolayer. Consequently, we can switch on and off the NDC and continuously tune the regions of the NDC in the spectra simply by using the magnetic field.

In summary, we realized magnetic-field-controlled NDC in graphene *npn* junction resonators. Electron confinement in the graphene *npn* junctions helps to generate sharp and pronounced LLs in the presence of large magnetic fields. The tip-induced electrostatic potential leads to misalignment between the sharp and pronounced LLs of the graphene region beneath the STM tip and that of the surrounding bulk. Tunneling between the misaligned LLs results in the observed strong NDC. Our experiment explicitly demonstrated that we can use the magnetic field to switch on and off the NDC and to continuously tune the regions of the NDC in the spectra. This result is not only very important in understanding exotic phenomena of the graphene *npn* junction resonators in the presence of magnetic fields, but also in potential applications of advanced nanoelectronics.

**Acknowledgments**

This work was supported by the National Natural Science Foundation of China (Grant Nos. 11674029, 11422430, 11374035, 11374219, 11504008, 11674028), the Natural Science Foundation of Jiangsu Province, China (Grant No. BK20160007), the National Basic Research Program of China (Grants Nos. 2014CB920903, 2013CBA01603, 2014CB920901), the program for New Century Excellent Talents in University of the Ministry of Education of China (Grant No. NCET-13-0054). L.H. also acknowledges support from the National Program for Support of Top-notch Young Professionals and support from "the Fundamental Research Funds for the Central Universities".

**Author contributions**

S.Y.L. performed the STM experiments and synthesized the samples. L.H., S.Y.L., and J.B.Q. analyzed the data. H.J. performed the theoretical calculations. L.H. conceived and provided advice on the experiment, analysis, and theoretical calculation. L.H. and S.Y.L. wrote the paper. All authors participated in the data discussion.

**Additional information**

Supplementary information is available in the online version of the paper. Reprints and permissions information is available online at www.nature.com/reprints. Correspondence and requests for materials should be addressed to H.J. or L.H.

**Reference**

[1] Katsnelson, M. I., Novoselov, K. S. & Geim, A. K. Chiral tunnelling and the Klein paradox in graphene. *Nature Phys.* **2**, 620 (2006).

[2] Pereira, J. M., Mlinar, V., Peeters, F. M. & Vasilopoulos, P. Confined states and direction-dependent transmission in graphene quantum wells. *Phys. Rev. B* **74**, 045424 (2006).

[3] Cheianov, V. V. & Fal'Ko, V. I. Selective transmission of Dirac electrons and ballistic magnetoresistance of n-p junctions in graphene. *Phys. Rev. B* **74**, 041403 (2006).

[4] Young, A. & Kim, P. Quantum transport and Klein tunneling in graphene heterojunctions. *Nature Phys*. **5**, 222 (2008).

[5] Stander, N., Huard, B. & Goldhabergordon, D. Evidence for Klein tunneling in graphene p-n junctions. *Phys. Rev. Lett.* **102**, 026807 (2009).

[6] He, W. Y., Chu, Z. D. & He, L. Chiral tunneling in a twisted graphene bilayer. *Phys. Rev. Lett.* **111**, 066803 (2013).

[7] Park, C.-H., Yang, L., Son, Y.-W., Cohen, M. L., Louie, S. G. Anisotropic behaviours of massless Dirac fermions in graphene under periodic potentials. *Nature Phys*. **4,** 213 (2008).

[8] Cheianov, V. V., Fal'ko, V. & Altshuler, B. L. The Focusing of Electron Flow and a Veselago Lens in Graphene p-n Junctions. *Science* **315**, 1252 (2007).

[9] Miao, F. *et al*. Phase-coherent transport in graphene quantum billiards. *Science* **317**, 1530 (2007).

[10] Zhang, F. M., He, Y. & Chen, X. Guided modes in graphene waveguides. *Appl. Phys. Lett.* **94**, 212105 (2009).

[11] Williams, J. R., Low, T., Lundstrom, M. S. & Marcus, C. M. Gate-controlled guiding of electrons in graphene. *Nature Nanotechnol.* **6**, 222 (2011).

[12] Rickhaus, P. *et al*. Ballistic interferences in suspended graphene. *Nature Commun.* **4**, 2342 (2013).

[13] Taychatanapat, T., Watanabe, K., Taniguchi, T. & Jarillo-Herrero, P. Electrically

tunable transverse magnetic focusing in graphene. *Nature Phys.* **9**, 225 (2013).

[14] Lee, G. H., Park, G. H. & Lee, H. J. Observation of negative refraction of Dirac fermions in graphene. *Nature Phys.* **11**, 925 (2015).

[15] Chen, S. *et al.* Electron optics with p-n junctions in ballistic graphene. *Science* **353**, 1522 (2016).

[16] Zhao, Y. *et al.* Creating and probing electron whispering-gallery modes in graphene. *Science* **348**, 672 (2015).

[17] Lee, J. *et al.* Imaging electrostatically confined Dirac fermions in graphene quantum dots. *Nature Phys.* **12**, 1032 (2016).

[18] Gutierrez, C., Brown, L., Kim, C.-J., Park, J. & Pasupathy, A. N. Klein tunnelling and electron trapping in nanometre-scale graphene quantum dots. *Nature Phys.* **12**, 1069 (2016).

[19] Bai, K. K., Qiao, J. B., Jiang, H., Liu, H. W. & He, L. Massless Dirac Fermions Trapping in a Quasi-one-dimensional npn Junction of a Continuous Graphene Monolayer. *Phys. Rev. B* **95**, 201406(R) (2017).

[20] Jiang, Y., et al. Tuning a circular p-n junction in graphene from quantum confinement to optical guiding. arXiv: 1705.07346. To appear in Nature Nano.

[21] Bai, K.-K., Zhou, J.-J., Wei, Y.-C., Qiao, J.-B., Liu, Y.-W., Liu, H.-W., Jiang, H., He, L. Generating atomically-sharp p-n junctions in graphene and testing quantum electron-optics in nanoscale. arXiv: 1705.10952.

[22] Ghahari, F., et al. An on/off Berry phase switch in circular graphene resonators. *Science* **356**, 845 (2017).

[23] Chang, L. L., Esaki, L., Tsu, R. Resonant tunneling in semiconductor double barriers. *Appl. Phys. Lett.* **24**, 593 (1974).

[24] Chen, J., Reed, M. A., Rawlett, A. M., Tour, J. M. Large on-off ratios and negative differential resistance in a molecular electronic device. *Science* **286**, 1550-1552 (1999).

[25] Lyo, I., Avouris, P. Negative differential resistance on the atomic scale: implications for atomic scale devices. *Science* **245**, 1369-1372 (1989).

[26] Galperin, M., Ratner, M. A., Nitzan, A., Troisi, A. Nuclear coupling and polarization in molecular transport junctions: beyond tunneling to function. *Science* **319**, 1056-1060 (2008).

[27] Berthe, M., Stiufiuc, R., Grandidier, B., Deresmes, D., Delerus, C., Stievenard, D. Probing the carrier capture rate of a single quantum level. *Science* **319**, 436-438 (2008).

[28] Rashidi, M., Taucer, M., Ozfidan, I., Lloyd, E., Koleini, M., Labidi, H., Pitters, J.

L., Maciejko, J., Wolkow, R. A. Time-resolved imaging of negative differential resistance on the atomic scale. *Phys. Rev. Lett.* **117**, 216805 (2016).

[29] Nandkishore, R., Levitov, L. Common-path interference and oscillatory Zener tunneling in bilayer graphene p-n junctions. *Proc Natl Acad Sci* (*USA*) **108**, 14021-14025 (2011)

[30] Kim, K. S., Kim, T., Walter, A. L., Seyller, T., Yeom, H. W., Rotenberg, E., Bostwick, A. Visualizing atomic-scale negative differential resistance in bilayer graphene. *Phys. Rev. Lett.* **110**, 036804 (2013).

[31] Britnell, L., Gorbachev, R. V., Geim, A. K., Ponomarenko, L. A., Mishchenko, A., Greenaway, M. T., Fromhold, T. M., Novoselov, K. S., Eaves, L. Resonant tunneling and negative differential conductance in graphene transistors. *Nature Commun.* **4**, 1794 (2013).

[32] Mishchenko, A., et al. Twist-controlled resonant tunneling in graphene/boron nitride/graphene heterostructures. *Nature Nano.* **9**, 808 (2014).

[33] Yan, W., Liu, M., Dou, R.-F., Meng, L., Feng, L., Chu, Z.-D., Zhang, Y., Liu, Z., Nie, J.-C., He, L. Angle-dependent van Hove singularities in a slightly twisted graphene bilayer. *Phys. Rev. Lett*. **109**, 126801 (2012).

[34] Yan, W., Meng, L., Liu, M., Qiao, J.-B., Chu, Z.-D., Dou, R.-F., Liu, Z., Nie, J.-C., Naugle, D. G., He, L. Angle-dependent van Hove singularities and their breakdown in twisted graphene bilayers. *Phys. Rev. B* **90**, 115402 (2014).

[35] Zhang, Y. *et al.* Scanning tunneling microscopy of the π magnetism of a single carbon vacancy in graphene. *Phys. Rev. Lett.* **117**, 166801 (2016).

[36] Luican, A., Li, G., Reina, A., Kong, J., Nair, R. R., Novoselov, K. S., Geim, A. K., Andrei, E. Y. Single-Layer Behavior and Its Breakdown in Twisted Graphene Layers. *Phys. Rev. Lett.* **106**, 126802 (2011).

[37] Yin, L.-J., Qiao, J.-B., Wang, W.-X., Zuo, W.-J., Yan, W., Xu, R., Dou, R.-F., Nie, J.-C., He, L. Landau quantization and Fermi velocity renormalization in twisted graphene bilayers. *Phys. Rev. B* **92**, 201408(R) (2015).

[38] Yin, L.-J., Li, S.-Y., Qiao, J.-B., Nie, J,-C., He, L. Landau quantization in graphene monolayer, Bernal bilayer, and Bernal trilayer on graphite surface. *Phys. Rev. B* **91**, 115405 (2015).

[39] Bai, K.-K., Wei, Y.-C., Qiao, J.-B., Li, S.-Y., Yan, W., Nie, J.-C., He, L. Detecting giant electron-hole asymmetry in a graphene monolayer generated by strain and

charged-defect scattering via Landau level spectroscopy. *Phys. Rev. B* **92**, 121405(R) (2015).

[40] Zhang, Y., Brar, V. W., Girit, C., Zettl, A., Crommie, M. F. Origin of spatial charge inhomogeneity in graphene. *Nature Phys.* **5**, 722 (2009).

[41] Jung, S., Rutter, G. M., Klimov, N. N., Newell, D. B., Calizo, I., Hight-Walker, A. R., Zhitenev, N. B., Stroscio, J. A. Evolution of microscopic localization in graphene in a magnetic field from scattering resonances to quantum dots. *Nature Phys.* **7**, 245 (2011).

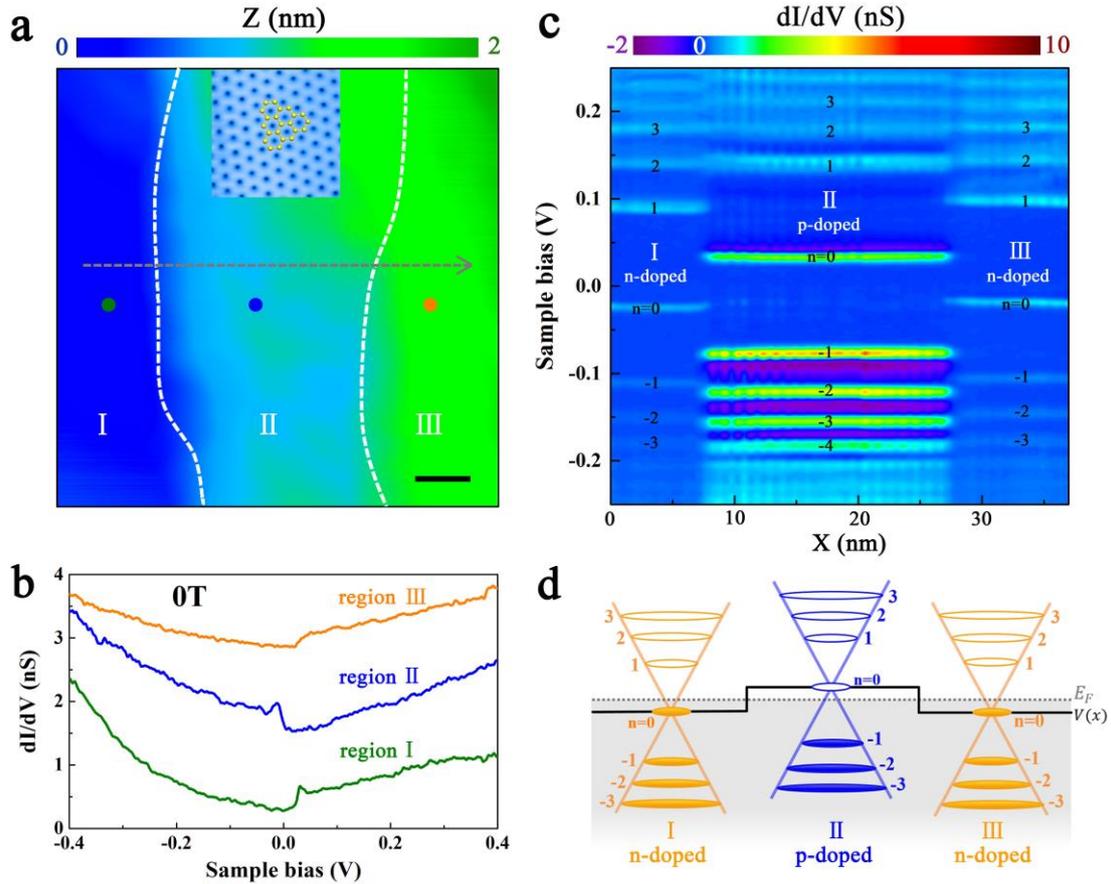

**Fig. 1. STM and STS measurements of a graphene *npn* junction resonator. a.** A 40 nm × 40 nm STM image of the graphene *npn* junction on Rh foil ($V_{sample}$ = 400 mV and $I$ = 0.3 nA). Scale bar: 5 nm. The color bar above shows the height fluctuation of this area and the two white dashed lines separate the studied area into three different regions. Inset: atomic-resolution STM image of the *p*-doped region showing hexagonal graphene lattices. **b.** The *dI/dV-V* spectra acquired from the three different regions of panel **a** (marked by solid circles with different colors) in zero magnetic field. **c.** STS map recorded along the grey dashed arrow in panel **a** in the presence of magnetic field 6 T. The LL indexes are marked. **d.** Schematic image for the *npn* junction of graphene monolayer in panel a under magnetic field 6 T. The grey dashed line marks the position of the Fermi level and the black solid line marks the potential barrier of the *npn* junction.

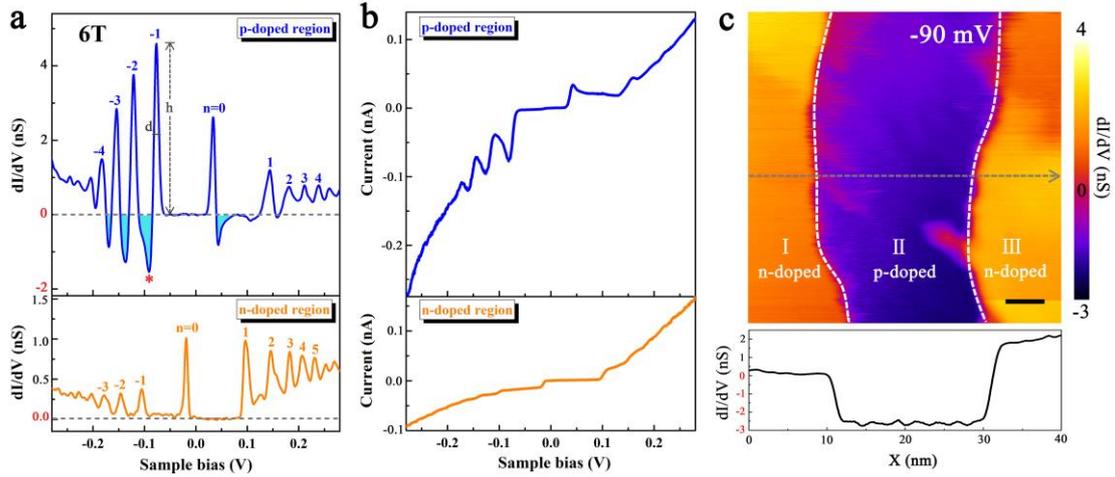

**Fig. 2. Strong NDC in the spectra recorded in the *p*-doped region. a.** The *dI/dV-V* spectra recorded in the *p*-doped region and the n-doped region in the presence of magnetic field of 6 T. The LL indexes are marked and the grey dashed lines mark the position of zero tunneling conductance. **b.** The corresponding *I-V* spectra acquired in the *p*-doped region and the *n*-doped region. **c.** Upper panel: STS map recorded in the *npn* junction of graphene monolayer at the fixed sample bias -90 mV ($I = 0.3$ nA) in the presence of magnetic field of 6 T. Scale bar: 5 nm. The white dashed lines mark the boundaries of the *p*-doped region in the *npn* junction. The color bar corresponds to the value of the differential tunneling conductance. Lower panel: the profile line across the upper STS map of the *npn* junction along the grey dashed arrow.

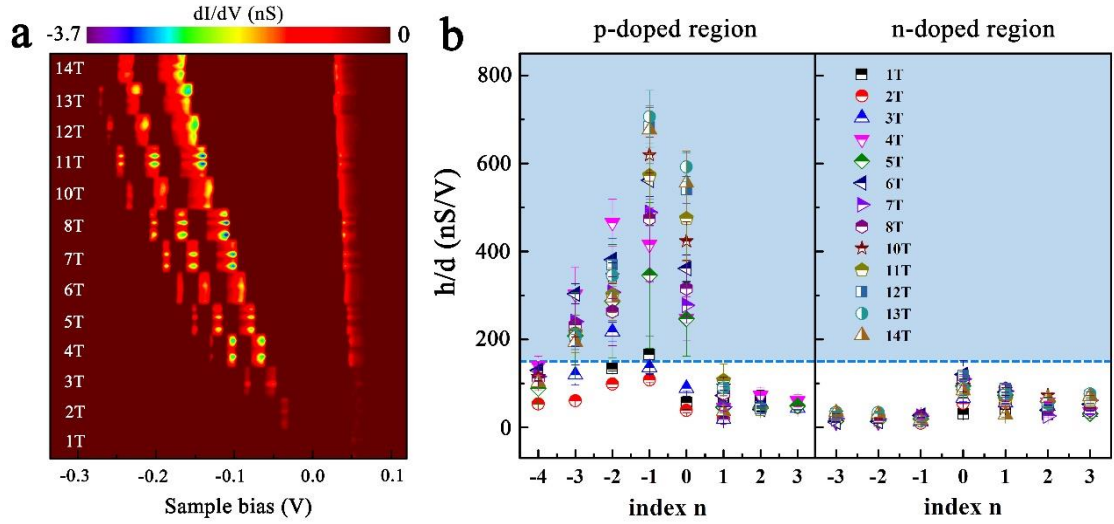

**FIG. 3. The observed NDC in the *p*-doped region as a function of magnetic field.**
**a.** Experimental result of the NDC recorded in the *p*-doped region as a function of magnetic field. Each panel of a fixed magnetic field contains ten typical spectra measured at different positions in the *p*-doped region. The horizontal axis for all panels is the sample bias voltage and the intensity of the NDC is shown in a color scale. Here we only show the tunneling conductance below zero of the STS spectra to emphasize the NDC. **b.** We summarize the ratios *h*/*d* for different LLs measured at different magnetic fields in the *p*-doped region and the *n*-doped region. Here *h* is the height of the LLs and *d* is the full width at half maximum of the LLs in the STS spectra, as defined in Fig. 2a. For those LLs with value of the ratio *h*/*d* above the dashed lines, we can observe the NDC accompanying them. This indicates that the sharp and pronounced LLs play a vital role in the emergence of the NDC.

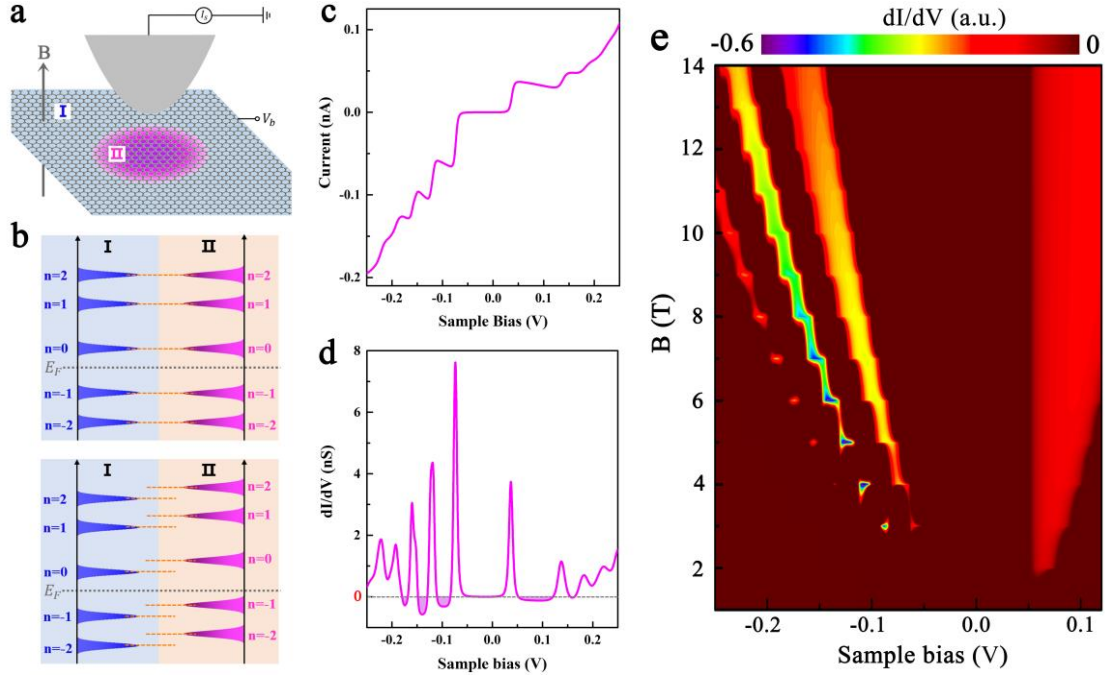

**FIG. 4. Theoretical calculation of the magnetic-field-controlled NDC in graphene monolayer. a.** Sketch of the experiment. The graphene monolayer is divided into two regions: one is the graphene region beneath the STM tip (region II), the other is the surrounding graphene region (region I). **b.** Schematic images showing relative positions of the LLs in the region I and region II. The Landau level indexes and the position of the Fermi level $E_F$ are marked. **c.** The theoretical *I-V* curve calculated according to Eq. (1) in the presence of magnetic field of 6 T. **d.** The corresponding theoretical *dI/dV-V* spectrum calculated according to Eq. (2) with the external magnetic field of 6 T. The grey dashed line marks the position of zero tunneling conductance. **e.** Theoretical calculated NDC as a function of magnetic field. The horizontal axis for all panels is the sample bias voltage and the intensity of the NDC is shown in a color scale.